\documentclass[12pt]{article}
\usepackage{amssymb}
\usepackage{epsfig}

\catcode`\@=11
\def\numberbysection{\@addtoreset{equation}{section}
        \def\theequation{\thesection.\arabic{equation}}}

\def\beq{\begin{equation}}
\def\eeq{\end{equation}}
\numberbysection
\begin{document}
\begin{titlepage}
\begin{center}
\hfill DFF  1/04/04 \\
\vskip 1.in {\Large \bf Projectors, matrix models and
noncommutative monopoles} \vskip 0.5in P. Valtancoli
\\[.2in]
{\em Dipartimento di Fisica, Polo Scientifico Universit\'a di Firenze \\
and INFN, Sezione di Firenze (Italy)\\
Via G. Sansone 1, 50019 Sesto Fiorentino, Italy}
\end{center}
\vskip .5in
\begin{abstract}
We study the interconnection between the finite projective modules
for a fuzzy sphere, determined in a previous paper, and the matrix
model approach, making clear the physical meaning of
noncommutative topological configurations.
\end{abstract}
\medskip
\end{titlepage}
\pagenumbering{arabic}
\section{Introduction}

One of the more interesting applications of noncommutative
geometry is the study of topologically nontrivial configurations
over a noncommutative manifold. In the commutative case, a general
method, based on the Serre-Swan's theorem allows to paraphrase the
study of nontrivial topologies for vector bundles in terms of
projective modules. At a classical level this theorem asserts the
complete equivalence between the category of vector bundles over a
compact manifold $M$ and the category of finite projective modules
over the commutative algebra $C(M)$ of ( smooth ) functions over
$M$.

The aim of this paper is to make an analogous bridge at a
noncommutative level, by studying an explicit example, i.e.
noncommutative monopoles on a fuzzy sphere \cite{a17}. In Ref.
\cite{a21} ( see also \cite{a18}-\cite{a20} ) we identified the
noncommutative projectors which should clarify the nontrivial
topologies on a fuzzy sphere, however to apply them in a physical
framework we need to deconstruct the projectors and to make a
bridge with the more familiar language of connections
\cite{a1}-\cite{a2}. On the other hand in Refs.
\cite{a22}-\cite{a23} we studied a matrix model, firstly
introduced in refs. \cite{a6}-\cite{a9} which defines $U(1)$
noncommutative gauge theory on a fuzzy sphere, based on a matrix
variable $X_i$ \cite{a10}-\cite{a16}, which contains the
information about the connection. We then investigated in Ref.
\cite{a23} the possible soliton solutions, but we didn't succeed
to identify the nontrivial topologies at the connection level.

Therefore our aim is to complete our knowledge on noncommutative
gauge theories on a fuzzy sphere by reaching mainly two
objectives: firstly we are able to identify the class of models
for which the projectors of Ref. \cite{a21} are indeed solution of
the $Y-M$ equations of motion, and secondly after deconstructing
the projectors in terms of more fundamental vector-valued fields,
we are able to identify the nontrivial connections.

The structure of the paper is as follows. Firstly we recall the
construction ( see \cite{a3} ) of projectors for a classical
monopole on $S^2$ by using the Hopf principal fibration $ \pi :
S^3 \rightarrow S^2 $ on the two-sphere, with $U(1)$ as a
structure group. Then we write the $Y-M$ equations of motion in
terms of a classical matrix model and show that the classical
projectors are solution to them. In section $4$ we recall the
construction of noncommutative projectors on a fuzzy sphere and,
using a simple relation between the matrix model variable $X_i$
and the noncommutative projectors we identify the class of models
which allow for nontrivial topologies as solutions to the $Y-M$
equations of motion. Finally we reconstruct the gauge connection
from the projectors and we give a nice interpretation of a
topologically nontrivial matrix variable $X_i$.

\section{Monopole solution in terms of projectors}

In Ref. \cite{a3} it has been shown how to characterize the
nontrivial connections on the two-sphere $S^2$ in terms of
projectors ( gauge invariant definition of a connection ), by
using the Hopf principal fibration $\pi : S^3 \rightarrow S^2$ on
the two-sphere $S^2$, with $U(1)$ as structure group.

The general procedure is starting from the algebra of $N \times N$
matrices whose entries are elements of the smooth function algebra
$ C^{\infty} ( S^2 ) $ on the base space $S^2$, i.e. $M_N (
C^{\infty} ( S^2 ) ) $. The section module of the bundle on which
the monopole lives can be identified with the action of a global
projector $ p \in M_N ( C^{\infty} ( S^2 ) ) $ on the trivial
module $ {( C^{\infty} ( S^2 ) ) }^N $, i.e. the right module $ p
{( C^{\infty} ( S^2 ) ) }^N $, where un element of $ {( C^{\infty}
( S^2 ) ) }^N $ is the vector

\beq  || f >> = \left( \begin{array}{c} f_1 \\ ... \\ f_N
\end{array}\right) \label{21}\eeq

with $ f_1, f_2, ...., f_N $ elements of $ {( C^{\infty} ( S^2 ) )
}^N $.

The projectors are written in general as  ket-bra valued functions
( with $N \times N$ entries )

\beq p = |\psi >< \psi |  \ \ \ \ \ < \psi | = < \psi_1 , ... ,
\psi_N | \label{22} . \eeq

The normalization condition on the vector valued function

\beq < \psi | \psi > = 1  \label{23}\eeq

automatically implies that $p$ is a projector since

\beq p^2 = |\psi >< \psi | \psi > < \psi | = p \ \ \ \ p^{\dagger}
= p \label{24}\eeq

and that the projector is of rank $1$ over $C$ since

\beq Tr (p) = < \psi | \psi > = 1 \label{25} . \eeq

We notice that it is possible to redefine the vector valued
function $ | \psi
> $  up to the right action of an element $ w \in U(1)
$, leaving the projector $p$ invariant.

The associated canonical connection is defined as :

\beq \nabla = p \cdot d \label{26}\eeq

and its curvature

\beq \nabla^2 = p ( dp )^2 = | \psi >< d \psi | d \psi >< \psi |
\label{27} . \eeq

The corresponding 1-form connection $A_\nabla$ has a very simple
expression in terms of the vector-valued function $ |\psi >$

\beq A_\nabla = < \psi | d \psi > \label{28} . \eeq

The classification of gauge non-equivalent connections depends on
the possible global left actions on the vector valued function
$|\psi
>$. If this global left action is reduced to the unitary group $SU(N) = \{
s | s s^{\dagger} = 1 \}$ preserving the normalization

\beq | \psi > \rightarrow | \psi^s > = s |\psi > \ \ \ \ \ \ <
\psi^s | \psi^s > = 1 \label{29}\eeq

then we remain in the same class of solutions since the connection
$A_\nabla$ is left invariant

\beq A^s_\nabla = < \psi^s | d \psi^s > = \psi | s^{\dagger} s d
\psi > =  A_\nabla \label{210} . \eeq

To obtain gauge non equivalent connections we should act with
group elements which do not preserve the normalization.

The explicit construction of nontrivial connections on the sphere
depicted in ref \cite{a3} involves the Hopf fibration. The $U(1)$
principal fibration $ \pi : S^3 \rightarrow S^2 $ is explicitly
realized as follows. The initial space is the $3$-dimensional
sphere:

\beq S^3 = \{ ( z_0, z_1 ) \in C^2 ; | z_0 |^2 + | z_1 |^2 = 1 \}
\label{211}\eeq

with right $U(1)$-action

\beq S^3 \times U(1) \rightarrow S^3 , \ \ \ \ \ ( z_0 , z_1
)\cdot w = ( z_0 w , z_1 w ) \label{212} . \eeq

The bundle projection $ \pi : S^3 \rightarrow S^2 $ defining the
Hopf fibration $ \pi ( z_0 , z_1 ) = ( x_1, x_2, x_3 ) $ is
determined to be

\begin{eqnarray}
x_1 & = & z_0 \overline{z}_1 + z_1 \overline{z}_0 \nonumber \\
x_2 & = & i ( z_0 \overline{z}_1 - z_1 \overline{z}_0 ) \nonumber
\\
x_3 & = & | z_0 |^2 - | z_1 |^2 \label{213} . \end{eqnarray}

We recall the projector construction for monopoles of charge $-n,
n \in N$. Let us consider the following vector-valued function
with $ N = n+1 $ components

\beq < \psi_{-n} | = \left( z_0 ^n , ...., \sqrt{ \left(
\begin{array}{c} n \\ k \end{array} \right) } z_0^{n-k} z_1^k, ....,
z^n_1 \right) \label{214}\eeq

where, for the general $n$-case we need to introduce the binomial
coefficients

\beq \left(
\begin{array}{c} n \\ k \end{array} \right) = \frac{ n! }{ k!
(n-k)! } \ \ \ \ \ \ \ \ k = 0,1,..n \label{215} . \eeq

Since the coordinates $ ( z_0, z_1)$ belong to the sphere $S^3$,
the normalization condition for the vector-valued function
(\ref{214}) is satisfied by construction :

\beq < \psi_{-n} | \psi_{-n} > = ( |z_0|^2 + |z_1|^2 )^n = 1
\label{216} . \eeq

From $< \psi_{-n} |$ we can deduce the projector

\beq p_{-n} = | \psi_{-n} >< \psi_{-n} | \label{217} . \eeq

The normalization condition (\ref{216}) ensures that $p_{-n}$ is a
projector

\begin{eqnarray}
p^2_{-n} & = & | \psi_{-n} >< \psi_{-n} | \psi_{-n} >< \psi_{-n} |
= | \psi_{-n} >< \psi_{-n} | = p_{-n} \nonumber \\
p^{\dagger}_{-n} & = & p_{-n} \label{218}\end{eqnarray}

and of rank $1$ since its trace is the constant $1$:

\beq Tr \ p_{-n} = < \psi_{-n} | \psi_{-n} > = 1 \label{219} .
\eeq

While $ < \psi_{-n} | $ is defined on the sphere $S^3$, the
projector $p_{-n}$ is defined on the sphere $S^2$, since the left
$U(1)$-action of $ < \psi_{-n} | $ leaves the projector $p_{-n}$
invariant:

\begin{eqnarray}
< \psi_{-n} | & \rightarrow & < \psi_{-n}^w | = w^n < \psi_{-n} |
\
\ \ \ \ \ \forall w \in U(1) \nonumber \\
p_{-n} & \rightarrow & p_{-n}^w = p_{-n} \ \ \ \ \ \overline{w} w
= 1 \label{220} . \end{eqnarray}

Therefore the projector entries can be expressed entirely as
functions of the base space $S^2$, as it should be.

To generalize the projector construction for monopoles of charge
$n$, we consider the vector-valued function:

\beq < \psi_n | = \left( \overline{z}_0 ^n , ...., \sqrt{ \left(
\begin{array}{c} n \\ k \end{array} \right) } \overline{z}_0^{n-k} \overline{z}_1^k, ....,
\overline{z}^n_1 \right) \label{221}\eeq

that is again normalized to 1.

The corresponding projector

\beq p_n = | \psi_n >< \psi_n | \label{222}\eeq

is invariant under the $U(1)$ action

\beq < \psi_n | \rightarrow < \psi_n^w | = \overline{w}^n < \psi_n
| \ \ \ \ \ \ \ \forall w \in U(1) \label{223} . \eeq

In this classical case since the functions $ < \psi_{-n} | $ and $
< \psi_n | $ are related by complex conjugation, the corresponding
projectors are related by hermiticity:

\beq p_n = {( p_{-n} )}^{\dagger} \label{224} . \eeq

The corresponding monopoles connections are proportional to the
charge number $n$ :

\begin{eqnarray}
A_{\mp n} & = & < \psi_{\mp n} | d \psi_{\mp n} > = \mp n (
\overline{z}_0 d z_0 + \overline{z}_1 d z_1 ) = \mp n A_1
\nonumber
\\
A_1 & = & \overline{z}_0 d z_0 + \overline{z}_1 d z_1 \label{225}
. \end{eqnarray}

The integer $n$ is related to the Chern number

\beq c_1 ( p_{\mp n} ) = - \frac{1}{2 \pi i} \int_{S^2} Tr (
\nabla^2_{\mp n} ) = - \frac{1}{2 \pi i} \int_{S^2} Tr ( p_{\mp n}
( dp_{\mp n})^2 ) = \pm n \label{226} . \eeq

\section{Classical Yang-Mills action and matrix models}

The monopole connection is not only a topological property of
$U(1)$ Yang-Mills theory on the two-sphere $S^2$ but it also
satisfies the equations of motion. Therefore our aim is rewriting
the Yang-Mills equations of motion in a form which makes manifest
that the projectors of the classical monopoles are solution to
them. This form will be used to make the extension to the
noncommutative case.

Based on our experience on the possible representations of the
Yang-Mills action, the most convenient choice turns out to rewrite
the classical Yang-Mills action in terms of matrix models and then
to connect the matrix model variables with the projectors.

Let us recall the most general $U(1)$ Yang-Mills action on the
sphere in terms of matrix models \cite{a6}-\cite{a9}. Let us
define the matrix variable

\beq X_i^{0} = L_i + A_ i \ \ \  \ \ \ \ \ [ L_i , . ] = - i k_i^a
\partial_a  \label{31}\eeq

where $A_i$ is related to the Yang-Mills connection $A_a (\Omega)$
and to an auxiliary scalar field $\phi ( \omega )$ as follows:

\beq A_i ( \Omega) = k_i^a A_a ( \Omega ) + \frac{x_i}{R} \phi(
\Omega) \ \ \ \ \ \ \Omega = ( \theta, \phi ) \label{32}\eeq

and $ k_i^a $ are the Killing vectors on the sphere $S^2$, and $R$
is its radius.

The classical action is defined as :

\begin{eqnarray}
S( \lambda ) & = & S_0 + \lambda S_1 = - \frac{1}{g^2_{YM}} \int d
\Omega \ \left[ \frac{1}{4} [ X^0_i, X^0_j ] [ X^0_i, X^0_j ] -
\frac{2}{3} i \lambda \epsilon^{ijk} X^0_i X^0_j X^0_k \right.
\nonumber
\\ & + & \left. ( 1 - \lambda ) X^0_i X^0_i \right]
\label{33} . \end{eqnarray}

Defining a gauge covariant field strength

\begin{eqnarray}
F_{ij} & = & [ X^0_i, X^0_j ] - i \epsilon^{ijk} X^0_k = \nonumber
\\ & = & [ L_i, A_j ] - [ L_j, A_i ] - i \epsilon_{ijk} A_k
\label{34}\end{eqnarray}

it can be developed in terms of the components fields (\ref{32})
as follows:

\beq F_{ij} ( \Omega ) =  k_i^a k_j^b F_{ab} + \frac{i}{R}
\epsilon_{ijk} x_k \phi - i \frac{x_j}{R} k^a_i
\partial_a \phi + i \frac{x_i}{R} k^a_j
\partial_a \phi \label{35}\eeq

where $F_{ab}= - i ( \partial_a A_b - \partial_b A_a ) $ for the
$U(1)$ case.

The classical Yang-Mills action on the sphere is thus
reconstructed:

\begin{eqnarray}
S( \lambda ) & = & - \frac{1}{4 g^2_{YM} } \int d \Omega [ (
F_{ab} + ( 4 - 2 \lambda ) i \epsilon_{ab} \phi \sqrt{g} )( F^{ab}
+ ( 4 - 2 \lambda ) i \epsilon^{ab} \frac{\phi }{\sqrt{g}} ) +
\nonumber \\
& - & 2 \partial_a \phi \partial^a \phi + ( 8 ( 2 - \lambda )^2 -
4 ( 2 - \lambda )) \phi^2 ] \label{36} . \end{eqnarray}

The auxiliary scalar field $\phi$ can be decoupled from pure
Yang-Mills theory for the particular value $\lambda = 2 $, in
which case the action reduces to :

\begin{eqnarray} S(2) & = & - \frac{1}{4 g^2_{YM} } \int d \Omega ( F_{ab}
F^{ab} - 2 \partial_a \phi \partial^a \phi ) \nonumber \\
d \Omega & = & \sqrt{g} d \theta d \phi = sin \theta d \theta d
\phi \ \ \ \ \ \ F^{ab} = g^{a a'} g^{b b'} F_{a' b'} \label{37} .
\end{eqnarray}

The equations of motion of the decoupled action $S(2)$ has two
basic solutions:

i) the monopole solution, with $ \phi = 0 $, which is the aim of
the present paper;

ii) $A_a = 0, \phi = const$, which has been considered in the
appendix of our paper \cite{a23}. In the following we will clarify
that a simple noncommutative map is able to connect the
noncommutative generalization of these two types of classical
solutions.

We notice that rewriting the classical action in terms of matrix
models, that becomes polynomial in the fundamental variables,
making simpler the link with the method of projectors.

It is easy to observe that the projector $p_0$ can be linked to
the matrix variables $X_i^0$ according to the following formula:

\beq \tilde{X}_i = p_0 L_i p_0 \label{38} . \eeq

Being the projector a gauge invariant formulation of the
connections, the matrix variable $\tilde{X}_i$ is again gauge
invariant and therefore it is not directly connected with $X_i^0$.
In any case shifting to the gauge covariant formulation is still
possible, since the projector $p_0$ can be put in the form of a
ket-bra valued function:

\beq p_0 = | \psi >< \psi | \label{39}\eeq

and , once the vector-valued function $ < \psi |$ is introduced,
the passage from $\tilde{X}_i$ to $X^0_i$ is manifest:

\beq X^0_i = < \psi | \tilde{X}_i | \psi > = < \psi | L_i | \psi >
= L_i + < \psi | [ L_i , | \psi > ] \label{310} . \eeq

As a byproduct we have obtained a representation of the connection
$A_i$ in terms of the vector-valued function $|\psi >$:

\beq A_i = < \psi | [ L_i, | \psi > ] \label{311} . \eeq

Before entering into details we firstly notice ( as in ref.
\cite{a1} ) that the direct introduction of the link (\ref{38})
between matrix model variables and projectors into the classical
action (\ref{33}) gives rise to a problem at a level of the
variational principle. In fact the equations of motion obtained by
varying the projector $p_0$, subject to the conditions $ p^2_0 =
p_0$, $ p^{\dagger}_0 = p_0 $, contain more solutions than those
obtained by varying the connection $X_i^0$.

To avoid such ambiguity we will limit ourself to introduce the
link with the projectors (\ref{38}) at the level of Yang-Mills
equations of motion and we will show that the monopole projectors
are indeed solutions of them.

By varying the classical Yang-Mills action (\ref{33}) with respect
to $X_i^0$ we find the following equation:

\beq [ X_j^0, F_{ij} ] = i ( \lambda - 1 ) \epsilon_{ijk} F_{jk}
\label{312}\eeq

where

\beq F_{ij} = [ X^0_i, X^0_j ] - i \epsilon_{ijk} X^0_k
\label{313} . \eeq

For $\lambda = 2 $ these matrix model equations of motion coincide
precisely with those of pure Yang-Mills theory on a two-sphere, by
simply posing the auxiliary scalar field $\phi = 0$.

We are ready to introduce the constraint $\tilde{X}_i = p_0 L_i
p_0 $ in (\ref{38}) and to verify that the monopole projectors
(\ref{217}) and (\ref{222})  are indeed solution of (\ref{312}).
Firstly we compute $F_{ij}$

\beq F_{ij} = p_0 ( [ L_i, p_0 ][ L_j, p_0 ] - [ L_j, p_0 ][ L_i,
p_0 ] ) \label{314}\eeq

where we made use of the property:

\beq p_0 [ L_i, p_0 ] p_0 = 0 \label{315} . \eeq

Therefore the equations (\ref{312}), rewritten in terms of a
generic projector $ p_0 $, take the following form:

\begin{eqnarray}
& \ & p_0 [ L^j, ( [ L_i, p_0 ][ L_j, p_0 ] - [ L_j, p_0 ][ L_i,
p_0 ] ) ] p_0 = \nonumber \\
& \ & = i ( \lambda -1 ) \epsilon_{ijk} p_0 ( [ L_i, p_0 ][ L_j,
p_0 ] - [ L_j, p_0 ][ L_i, p_0 ] ) p_0 \label{316} .
\end{eqnarray}

The evaluation of these equations on the classical projectors
(\ref{217}) and (\ref{222}) is surprisingly more difficult than
the noncommutative case as we shall see later. However we can
verify directly (\ref{316}) for the simplest case, the monopole
with charge $1$:

\beq p_1 = \frac{1}{2} \left(  \begin{array}{cc} 1+ cos \theta &
 sin \theta \ e^{- i \phi} \\ sin \theta \ e^{ i \phi }  & 1 - cos \theta \end{array}
  \right) \label{317} . \eeq

After simple calculations we get

\begin{eqnarray}
F_{\theta\phi} & = & p_1 ( \partial_\theta p_1 \partial_\phi p_1 -
\partial_\phi p_1 \partial_\theta p_1 ) = \frac{i}{2} sin \theta
\ p_1 \nonumber \\
F_{ij} & = & - ( k_i^\theta k_j^\phi - k_i^\phi k_j^\theta )
F_{\theta\phi} = - \frac{i}{2} p_1 \epsilon_{ijk} \frac{ x^k }{R}
\label{318} . \end{eqnarray}

It is then enough to apply the formula

\beq [ L_j , x_k ] = i \epsilon_{jkl} x^l \label{319}\eeq

and eq. (\ref{315}) to verify that $p_1$ is indeed solution of the
equations of motion (\ref{316}) for $\lambda = 2$.

Once clarified the interconnection between classical projectors
and Y-M equations, it is interesting to reconstruct the monopole
connections starting from the knowledge of the projectors $p_{\pm
n}$.

To reach this goal it is convenient to recall the formula
(\ref{311})

\beq A_i = < \psi | [ L_i , | \psi > ] \label{320}\eeq

where the commutator action reduces to a derivative action in the
classical case:

\begin{eqnarray} [ L_i, . ] &  = & -i ( k_i^\theta \partial_\theta + k_i^\phi
\partial_\phi + k_i^\psi \partial_\psi ) \nonumber \\
k_i^\theta & = & ( - sin \phi , cos \phi, 0 ) \nonumber \\
k_i^\phi & = & ( - cos \phi cotg \theta , - sin \phi cotg \theta , 1 ) \nonumber \\
k_i^\psi & = & ( - \frac{cos \phi}{sin \theta} , - \frac{ sin
\phi}{sin\theta}, 0 ) \label{321}\end{eqnarray}

where $k_i^\theta, k_i^\phi$ are the killing vector of the sphere
and $k_i^\psi$ is the residue of the extended action on $S^3$.
Unfortunately the derivative action contains the dependence on the
auxiliary $U(1)$ variable $\psi$ of the three sphere $S^3$, since
the vector-valued function $ < \psi |$, by construction, depends
on the total space $S^3$, while the projectors depend only on the
physical base space $S^2$.

Therefore for the simplest monopole $n=1$

\beq < \psi_1 | = ( \overline{z}_0, \overline{z}_1 ) \label{322}
\eeq

with

\beq z_0 = cos \frac{\theta}{2} e^{i \left( \frac{\psi-\phi}{2}
\right) } \ \ \ \ \ z_1 = sin \frac{\theta}{2} e^{i \left(
\frac{\psi+\phi}{2} \right) } \label{323}\eeq

we obtain at first sight for the connection

\begin{eqnarray}
A_i & = & k_i^\theta A_\theta + k_i^\phi A_\phi + k_i^\psi A_\psi
\nonumber \\
A_\psi & = & \frac{1}{2} \ \ \ \ \ A_\theta = 0 \ \ \ \ \ A_\phi =
- \frac{1}{2} cos \theta \label{324} .
\end{eqnarray}

We haven't yet been able to identify the monopole connection for
the presence of the spurious component $A_\psi$. However since the
field strength $F_{ij}$ depends only on the variables $\theta$ and
$\phi$, being determined by the projectors $p_{\pm n}$ as in the
formula (\ref{314}), it turns out that the presence of the
component $A_\psi$ is purely fictitious and can be removed by
redefining the vector-valued function $ < \psi_1 |$:

\beq < \psi_1 | \rightarrow < \psi'_1 | = e^{i\frac{\psi}{2}} <
\psi_1 | \ \ \ \ \ {\rm for \ } n = 1 \label{325}\eeq

or in the general case

\beq < \psi_n | \rightarrow < \psi'_n | = e^{i n \frac{\psi}{2}} <
\psi_n | \ \ \ \ \ \forall \ n \in N  \label{326} . \eeq

After this redefinition, which doesn't alter the projectors, we
obtain the well-known result for the monopole connection:

\beq A_\theta = 0 \ \ \ A_\phi = -\frac{n}{2} cos \theta
\label{327} . \eeq

We have therefore learned that the Lie derivative action is
equivalent to the ordinary derivate action $ d = d\theta
\frac{\partial}{\partial \theta} + d\phi \frac{\partial}{\partial
\phi} $ if and only if the vector-valued function $ < \psi_n |$
can be projected to the basic space $S^2$. This redefinition,
which is simple in the classical case, turns out to be necessary
also in the noncommutative case.

\section{Noncommutative projectors}

In Ref. \cite{a21} ( see also \cite{a18}-\cite{a20} ) , we have
been able to extend the classical monopoles in terms of new
noncommutative projectors having as entries the elements of the
fuzzy-sphere algebra:

\begin{eqnarray}
& \ & [ \hat{x}_i, \hat{x}_j ] = i \alpha \epsilon_{ijk} \hat{x}_k
\ \ \ \ \ \sum_i {( \hat{x}_i )}^2 = R^2 \nonumber \\
& \ & \alpha = \frac{2 R}{\sqrt{N ( N+2 )}} \label{41} .
\end{eqnarray}

In the $N\rightarrow \infty$ limit this algebra reduces to the
classical two-sphere. Our construction is based on the observation
that the classical $x_i$ coordinates are related to a couple of
complex coordinates by the Hopf principal fibration. The
noncommutativity between the coordinates is then realized by
promoting the two complex coordinates to a couple of independent
oscillators:

\begin{eqnarray}
z_i & \rightarrow & a_i \ \ \ \ \ \  [ a_i, a^{\dagger}_j ] =
\delta_{ij} \nonumber \\
\hat{x}_1 & = & \frac{\hat{\alpha}}{2} ( a_0 a_1^{\dagger} + a_1
a_0^{\dagger} ) \nonumber \\
\hat{x}_2 & = & i \frac{\hat{\alpha}}{2} ( a_0 a_1^{\dagger} - a_1
a_0^{\dagger} ) \nonumber \\
\hat{x}_3 & = & \frac{\hat{\alpha}}{2} ( a_0 a_0^{\dagger} - a_1
a_1^{\dagger} ) \nonumber \\
\hat{N} & = & a^{\dagger}_0 a_0 + a^{\dagger}_1 a_1 \label{42} .
\end{eqnarray}

The operator $\hat{\alpha}$ is equal to $\alpha$ for
representations with fixed total oscillator number $\hat{N} = N$,
which is the characteristic noncommutative parameter of the fuzzy
sphere. As it happens in the principal Hopf fibration we can
redefine the oscillators with a $U(1)$ factor which is cancelled
in the combinations $\hat{x}_i$.

Basically, the projectors are constructed in terms of
vector-valued operators $ < \psi | $, taking values in the
oscillator algebra, while the projectors will be dependent on
polynomial functions of the fuzzy-sphere algebra $\hat{x}_i$ only.

To construct the projectors $p_n ( \hat{x}_i)$ we have considered
the $ ( n+1) $-dimensional vectors:

\beq | \psi_n > = N_n \left( \begin{array}{c} (a_0)^n \\
..... \\ \sqrt{\left( \begin{array}{c} n \\ k \end{array} \right)}
(a_0)^{n-k} (a_1)^k \\ ...... \\ ( a_1 )^n
\end{array} \right)
\label{43} . \eeq

Constraining the vector $ < \psi_n | $ to be normalized one notice
that the function $ N_n $ is fixed to be dependent only on the
number operator $ \hat{N} $

\begin{eqnarray}
& \ & < \psi_n | \psi_n > = 1 \nonumber \\
& \ & N_n = N_n ( \hat{N} ) = \frac{1}{ \sqrt{\prod^{n-1}_{i=0} (
\hat{N} - i + n )} } \label{44} . \end{eqnarray}

Then the $n$-monopoles projector is simply

\beq p_n = | \psi_n >< \psi_n | \label{45}\eeq

and it satisfies the basic properties of a projector, due to the
normalization condition (\ref{44}).

It is easy to notice that in the ket-bra product there appear only
combinations of oscillators, commuting with the number operator $
\hat{N} $, and therefore the action of $ p_n $ can be restricted
to a fixed value $ \hat{N} = N $ and its entries belong to the
fuzzy-sphere algebra.

Moreover the projector $P_n$ has a positive trace given by

\beq Tr \  p_n = Tr \ | \psi_n >< \psi_n | = \frac{ N+n+1}{ N+1 }
\ Tr I = N + n + 1 < \ Tr \ I = ( N+1 ) ( n+1 ) \label{46} . \eeq

Being the algebra of the fuzzy sphere a finite-dimensional
algebra, the trace of the projector $p_n$ is always a positive
integer and less than the trace of the identity.

To construct the solution for $p_{-n} ( n > 0)$ it is enough to
take the adjoint of the components of the vector (\ref{43}), apart
the normalization factor, which is different in this case.
Consider the $(n+1)$-dimensional vectors:

\beq | \psi_{-n} > = N_n \left( \begin{array}{c} (a_0^\dagger)^n \\
..... \\ \sqrt{\left( \begin{array}{c} n \\ k \end{array} \right)}
(a_0^\dagger)^{n-k} (a_1^\dagger)^k \\ ...... \\ ( a_1^\dagger )^n
\end{array} \right) \label{47} .
\eeq

Again normalizing the vector $ < \psi_{-n} | $ fixes the function
$N_n$ to be dependent only on the number operator:

\begin{eqnarray}
& \ & < \psi_{-n} | \psi_{-n} > = 1 \nonumber \\
& \ & N_n = N_n ( \hat{N} ) = \frac{1}{ \sqrt{\prod^{n-1}_{i=0} (
\hat{N} + i + 2 - n )} } \label{48} .
\end{eqnarray}

The corresponding projector $p_{-n}$ doesn't exist for all values
of $n$, but only for $ 0 < n < N+2$. In fact the corresponding
trace is given by:

\beq Tr \ p_{-n} = Tr \ | \psi_{-n} >< \psi_{-n} | = \frac{ N+1-n
}{ N+1 } \ Tr \ I = N+1-n < \ Tr \ I \label{49} . \eeq

This trace is an integer, but it is positive definite if and only
if the following bound is respected:

\beq n < N+1 \label{410} . \eeq

For the special case $n = N+1$ we simply obtain a null projector.

In summary the projectors $p_{\pm n}( \hat{x}_i)$ have the nice
property to be a smooth deformation of the classical projectors
$p_{\pm n}( x_i)$, making evident the existence of noncommutative
monopoles which tends for $N \rightarrow \infty$ to the classical
ones. We recall that this procedure has been successfully extended
to the fuzzy four-sphere case \cite{a24}.

To construct the corresponding connection we need more work
because we must be sure that the action of the Lie derivative $[
L_i, . ] $ on the vectors $|\psi_{\pm n }>$ has as a smooth limit
the classical Lie derivative on the sphere and this criterium
requires an adjustment of the present construction.

\section{Properties of noncommutative projectors}

Let us start from the generic noncommutative projectors $p_n$;
defining the corresponding matrix model variables as in eq.
(\ref{38}), we easily obtain, using the oscillator algebra,
\footnote{For the simplest case $n=1, N=1$ we have checked that
the result can be expressed as a sum of two independent angular
momenta, one is intrinsic to the fuzzy sphere algebra $L_i \otimes
1$ and the other $1 \otimes S_i$ acting on the  $(n+1) \otimes
(n+1)$ auxiliary representation space of the projector.}

\begin{eqnarray}  \tilde{X}_i^{(n)} & = & p_n L_i p_n = \frac{ N }{ N+n } | \psi_n >
L_i < \psi_n | = | \psi_n > L_i < \psi_n | + | \psi_n > A_i^{(n)}
< \psi_n |
\ \  n > 0 \nonumber \\
 \tilde{X}_i^{(-n)} & = & p_{(-n)} L_i p_{(-n)} = \frac{ N+2 }{ N+2-n } | \psi_{(-n)} >
L_i < \psi_{(-n)} | = | \psi_{(-n)} > L_i < \psi_{(-n)} | +
\nonumber \\
&  \  &  + \ | \psi_{(-n)} > A_i^{(-n)} < \psi_{(-n)} | \ \ \ \ \
0 < n < N+1 \label{51} . \end{eqnarray}

The ( gauge invariant ) connection is not only proportional to the
classical monopole charge $n$ but it has another dependence from
$n$ in the denominator:

\begin{eqnarray} | \psi_n > A_i^{(n)} < \psi_n | & = & - \frac{ n }{ N+n } | \psi_n > L_i <
\psi_n | \ \ n > 0 \nonumber \\ | \psi_{(-n)} > A_i^{(-n)} <
\psi_{(-n)} | & = & \frac{ n }{ N+2-n } | \psi_{(-n)} > L_i <
\psi_{(-n)} | \nonumber \\ & \ & 0 < n < N+1 \label{52} .
\end{eqnarray}

Let's see what happens for the field strength:

\begin{eqnarray} \tilde{F}_{ij}^{(n)} & = & - \frac{n N}{(N+n)^2} i
\epsilon_{ijk} | \psi_n
> L_k < \psi_n | \ \ \ n > 0 \nonumber \\
\tilde{F}_{ij}^{(-n)} & = &  \frac{n(N+2)}{(N+2-n)^2}  i
\epsilon_{ijk} |
\psi_{(-n)} > L_k < \psi_{(-n)} | \nonumber \\
& \ & \ \ \ 0 < n < N+1 \label{53} .
\end{eqnarray}

Differently from the classical case, the field strength is no more
simply proportional to the instanton number $n$ unless in the $N
\rightarrow \infty$ limit; this is the main obstacle to define an
integer Chern class for the noncommutative monopole, due to the
nonlinear additional contribution.

The corresponding equations of motion are solved by a certain
value of $\lambda$, due to the identity:

\begin{eqnarray}
& \ & [ \tilde{X}^{(n)}_j, \tilde{F}^{(n)}_{ij} ] = i
\frac{N}{N+n} \epsilon_{ijk} \tilde{F}^{(n)}_{jk} \ \ \ \  \
\lambda
= 2 - \frac{n}{N+n} \ \ n > 0 \nonumber \\
& \ & [ \tilde{X}^{(-n)}_j, \tilde{F}^{(-n)}_{ij} ] = i
\frac{N+2}{N+2-n} \epsilon_{ijk} \tilde{F}^{(-n)}_{jk} \ \ \ \  \
\lambda
= 2 + \frac{n}{N+2-n} \ \ 0 < n < N+1 \nonumber \\
 \label{54} .
\end{eqnarray}

We see therefore that the class of models we are interested in is
situated around the classical value $\lambda = 2$ confirming an
interpretation already outlined in our paper \cite{a23}. In
particular we noticed that to define noncommutative soliton
solutions, having a smooth limit to the classical ones, it was
necessary to perturb the $\lambda$ coupling constant in a similar
form:

\beq \lambda = \lambda_{cl} + \frac{c}{N} \label{55} . \eeq

For the special case $ \lambda_{cl} = 2$ we already found a class
of solutions tending to the classical solution $ \phi = {\rm
const.}, A_a = 0 $; in this case the matrix model variable $X_i$
was equal to a rescaling of the background solution by a factor $(
1 + f(\frac{1}{N}))$:

\beq X_i = ( 1 + f(\frac{1}{N})) \hat{x}_i \ \ \ \rightarrow \ \ \
\phi = { \rm const } \ \ \ A_a = 0 \label{56} . \eeq

Evidently, inside the class of models (\ref{54}) it is possible to
reach a different classical limit:

\beq X_i \rightarrow \phi = 0 , \ \ A_a =  {\rm monopole }
\label{57}\eeq

and we will discuss later how this fact happens. We have already
reached an important result, i.e. we know for what class of models
our noncommutative projectors are solutions of the YM equations of
motion.

\section{Reconstruction of the gauge connection from projectors}

The reconstruction of the gauge connection from the noncommutative
projectors requires another step, since the vector $ < \psi_n |$,
used for the computation in (\ref{43}), is function of the
oscillator algebra, which is more general of the fuzzy sphere
algebra. This produces a discontinuity problem in the classical
limit. In fact the Lie derivative, when it acts on the oscillator
algebra, has the form

\begin{eqnarray}
L_x & = & \frac{1}{2} [ a_0 a_1^{\dagger} + a_1 a_0^{\dagger}, . ]
\nonumber
\\
L_y & = & \frac{1}{2} [ i ( a_0 a_1^{\dagger} - a_1
a_0^{\dagger}), . ] \nonumber
\\
L_z & = & \frac{1}{2} [ a_0 a_0^{\dagger} - a_1 a_1^{\dagger}, . ]
\label{61}
\end{eqnarray}

and it can be represented in the classical limit with the extended
action:

\begin{eqnarray}
L_i & = & -i ( k_i^\theta \partial_\theta + k_i^\phi \partial_\phi
+ k_i^\psi \partial_\psi ) \nonumber \\
k_i^\theta & = & ( - sin \phi , cos \phi, 0 ) \nonumber \\
k_i^\phi & = & ( - cos \phi cotg \theta , - sin \phi cotg \theta , 1 ) \nonumber \\
k_i^\psi & = & ( - \frac{cos \phi}{sin \theta} , - \frac{ sin
\phi}{sin\theta}, 0 ) \label{62} .
\end{eqnarray}

As a consequence, a spurious dependence on the variable $\psi \in
S^3$ is generated, which is physically irrelevant since the field
strength is function only of the two-sphere $S^2$.

At the classical level we already discussed such problem and we
have observed that the vector $ |\psi >$ has a sort of $U(1)$
gauge arbitrariness $ | \psi > \rightarrow | \psi > e^{i\phi}$
leaving the projector invariant.

What is the analogue of this phase ambiguity at a noncommutative
level ? Let's write for example the vector-valued operator $| \psi
>$ in the case $ n = - 1$:

\begin{eqnarray}
| \psi_{n=-1} > & = & \frac{1}{\sqrt{ \hat{N} + 1}} \left(
\begin{array}{c} \overline{a}_0 \\ \overline{a}_1 \end{array}
\right) = \left(
\begin{array}{c} \tilde{\overline{a}}_0 \\ \tilde{\overline{a}}_1 \end{array}
\right) \nonumber \\
\tilde{\overline{a}}_0 & = & \sum_{n_1 = 0}^{\infty} \sum_{n_2 =
0}^{\infty} \sqrt{ \frac{ n_1 +1}{n_1+n_2+2}} | n_1 +1, n_2 ><
n_1, n_2 | \nonumber \\
\tilde{\overline{a}}_1 & = & \sum_{n_1 = 0}^{\infty} \sum_{n_2 =
0}^{\infty} \sqrt{ \frac{ n_2 +1}{n_1+n_2+2}} | n_1, n_2 + 1><
n_1, n_2 | \nonumber \\
< \psi_{n=-1} | & = &  ( a_0 , a_1 ) \frac{1}{\sqrt{ \hat{N} + 1}}
= ( \tilde{a}_0 ,  \tilde{a}_1 ) \nonumber \\
\tilde{a}_0 & = & \sum_{n_1 = 0}^{\infty} \sum_{n_2 = 0}^{\infty}
\sqrt{ \frac{ n_1 +1}{n_1+n_2+2}} | n_1 , n_2 ><
n_1 + 1, n_2 | \nonumber \\
\tilde{a}_1 & = & \sum_{n_1 = 0}^{\infty} \sum_{n_2 = 0}^{\infty}
\sqrt{ \frac{ n_2 +1}{n_1+n_2+2}} | n_1, n_2 >< n_1, n_2 + 1 |
\nonumber \\
< \psi_{-1} | \psi_{-1} > & = & 1 \label{63} .
\end{eqnarray}

The problem which complicates the classical limit is that the
action of $ | \psi_{-1} >$ doesn't commute with the number
operator $\hat{N}$ and therefore it is not possible to restrict
its action to a fixed number $N$, as instead we have done for the
projectors. It is then necessary to correct the vector $ |
\psi_{-1}>$ with an operator, acting on the right and not
commuting with the number operator, i.e. a quasi-unitary operator:

\beq |\psi_{-1} > \rightarrow | \psi'_{-1}> = | \psi_{-1} > U \ \
\ \ \ \ \ U U^\dagger = 1 \ \ \ \ \ ( U^\dagger U = 1 - P_0 )
\label{64}\eeq

in order to keep invariant the noncommutative projectors. The
presence of the quasi-unitary operator $U$ adjusts the classical
limit, making possible to extrapolate the noncommutative
connection. Many choices for $U$ are possible, for example:

\beq U_1 = \sum_{n_1 = 0}^{\infty} \sum_{n_2 = 0}^{\infty}  | n_1,
n_2 >< n_1 + 1 , n_2  | \label{65}\eeq

or

\beq U_2 = \sum_{n_1 = 0}^{\infty} \sum_{n_2 = 0}^{\infty}  | n_1,
n_2 >< n_1 , n_2 +1 | \label{66}\eeq

are equally good, since the difference between these two operators
is a gauge transformation of the connection. Let's compute

\begin{eqnarray}
| \psi'_{-1} > & = & | \psi_{-1} > U_1 = \left(
\begin{array}{c} \tilde{\overline{a'}}_0 \\ \tilde{\overline{a'}}_1 \end{array}
\right) \nonumber \\
\tilde{\overline{a'}}_0 & = & \sum_{n_1 = 0}^{\infty} \sum_{n_2 =
0}^{\infty} \sqrt{ \frac{ n_1 +1}{n_1+n_2+2}} | n_1 +1, n_2 ><
n_1+1, n_2 | \nonumber \\
\tilde{\overline{a'}}_1 & = & \sum_{n_1 = 0}^{\infty} \sum_{n_2 =
0}^{\infty} \sqrt{ \frac{ n_2 +1}{n_1+n_2+2}} | n_1, n_2 + 1><
n_1+1, n_2 | \label{67} .
\end{eqnarray}

At this point $ [ \hat{N}, | \psi'_{-1} > ] = 0$ and we can
truncate the action of $|\psi'_{-1}>$ to a fixed number $N$:

\begin{eqnarray}
\tilde{\overline{a'}}_0 |_N & = & \sum_{k = 0}^{N-1}  \sqrt{
\frac{ k + 1}{N+1}} | k+1 , N-k-1 ><
k+1, N-k-1 | \nonumber \\
\tilde{\overline{a'}}_1 |_N & = & \sum_{k = 0}^{N-1}  \sqrt{
\frac{N-k}{N+1}} | k, N-k >< k+1, N-k-1 | \label{68} .
\end{eqnarray}

These actions can be reexpressed in terms of spherical harmonics
of the fuzzy sphere, i.e. the physical functional space of the
noncommutative case.

We can verify that $| \psi'_{-1} >$ gives rise to connections
satisfying the Y-M equations of motion ( in the gauge-covariant
formulation ). Since

\beq \tilde{X}_i = | \psi_{-1} > U \frac{N+2}{N+1} ( U^{\dagger}
L_i U ) U^{\dagger} < \psi_{-1} | = | \psi'_{-1} > X_i < \psi_{-1}
| \label{69} \eeq

we can deduce that the physical matrix model variable $X_i$ is of
the form:

\begin{eqnarray}
& \ & X_i = \frac{N+2}{N+1} U^{\dagger} L_i U \nonumber \\
& \ & F_{ij} = [ X_i, X_j ] - i \epsilon_{ijk} X_k =
\frac{N+2}{(N+1)^2} i
\epsilon_{ijk} ( U^{\dagger} L_j U ) \nonumber \\
& \ & [ X^j, F_{ij} ]  =  i ( \lambda - 1 ) \epsilon_{ijk} F_{jk}
\ \ \ \ \ \lambda = 2 + \frac{1}{N+1} \label{610} .
\end{eqnarray}

The generalization of these results to the case $|\psi_{-n}>$ with
$n$ generic ( $n < N+1$ ) is straightforward.

We can summarize these results as follows. The matrix model
solution $X_i$ for charge $-n$ is obtained in two steps:

i) re-scaling  the background solution $ X^{(0)}_i =
\frac{N+2}{N+2-n} L_i $, as already noticed in the appendix of the
paper \cite{a23}, leading to the classical solution $ \phi =
const., A_a = 0$;

ii) dressing with the quasi-unitary operator $X_i = U^\dagger
X^{(0)}_i U$, shifting the solution in another class with respect
to i), the monopole class. Therefore the noncommutative map
between the two classes is realized with a quasi-unitary operator

\beq U : ( \phi = {\rm const.} , A_a = 0 ) \rightarrow ( \phi = 0,
A_a = {\rm monopole } ) \label{611} . \eeq

We note that the action of quasi-unitary operators on the
background generates reducible representations of the $SU(2)$ Lie
algebra, revealing their topological character. Therefore the
classification of nontrivial topologies on a fuzzy sphere is
reduced to the classification of reducible representations of the
$SU(2)$ Lie algebra \footnote{Since there are no one-dimensional
representations of $SU(2)$ Lie algebra, the case $n=N$ leads to a
vanishing matrix variable, also if the corresponding projector is
non-vanishing.}.

It remains to be investigated if this construction can be repeated
for the case $|\psi_{n}>, \ \ n >0$, for example $n=1$:

\begin{eqnarray}
| \psi_{n=1} > & = & \frac{1}{\sqrt{ \hat{N} + 1}} \left(
\begin{array}{c} a_0 \\ a_1 \end{array}
\right) = \left(
\begin{array}{c} \tilde{a}_0 \\ \tilde{a}_1 \end{array}
\right) \nonumber \\
\tilde{a}_0 & = & \sum_{n_1 = 0}^{\infty} \sum_{n_2 = 0}^{\infty}
\sqrt{ \frac{ n_1 +1}{n_1+n_2+1}} | n_1, n_2 ><
n_1+1, n_2 | \nonumber \\
\tilde{a}_1 & = & \sum_{n_1 = 0}^{\infty} \sum_{n_2 = 0}^{\infty}
\sqrt{ \frac{ n_2 +1}{n_1+n_2+1}} | n_1, n_2 ><
n_1, n_2 +1| \nonumber \\
< \psi_{n=1} | & = & \frac{1}{\sqrt{ \hat{N} + 1}} (
\overline{a}_0 , \overline{a}_1 )
= ( \tilde{\overline{a}}_0 ,  \tilde{\overline{a}}_1 ) \nonumber \\
\tilde{\overline{a}}_0 & = & \sum_{n_1 = 0}^{\infty} \sum_{n_2 =
0}^{\infty} \sqrt{ \frac{ n_1 +1}{n_1+n_2+1}} | n_1 +1 , n_2 ><
n_1, n_2 | \nonumber \\
\tilde{\overline{a}}_1 & = & \sum_{n_1 = 0}^{\infty} \sum_{n_2 =
0}^{\infty} \sqrt{ \frac{ n_2 +1}{n_1+n_2+1}} | n_1, n_2 +1 ><
n_1, n_2 |
\nonumber \\
< \psi_{1} | \psi_{1} > & = & 1  - | 0,0 >< 0,0| = 1 - P_0
\label{612} .
\end{eqnarray}

The last normalization condition is equivalent to the identity
since:

\beq | \psi_1 > P_0 = P_0 < \psi_1 | = 0 \label{613}\eeq

the action of $ | \psi_{1} >$ on the projector $P_0$ is null.

If we try to redefine $ | \psi_1>$  in order to commute with the
number operator, we are forced to introduce the adjoint of the
quasi-unitary operator $U$

\beq U^{\dagger} = \sum_{n_1 = 0}^{\infty} \sum_{n_2 = 0}^{\infty}
 | n_1+1 , n_2  >< n_1, n_2 | \label{614} . \eeq

 However in this case the dressing unfortunately alters the form
 of the projector $p_1$ since

 \beq | \psi'_{1} >< \psi'_{1} | = | \psi_1 > U^{\dagger} U < \psi_1 |
 = | \psi_1 > ( 1 - \sum_{n=0}^{\infty} | 0, n >< 0, n| ) < \psi_1
 | \label{615} . \eeq

 In fact the extra contribution is not cancelled by the presence
 of $ | \psi_1 >$. We have checked, for the special case $N=1$
 using the basis of Pauli matrices, that only the combination
 $1-p_1$ can be written as a ket-bra valued operator, but
 unfortunately $1-p_1$ doesn't satisfy the Y-M equations of
 motion. We therefore find a contradictory result,
 since it is impossible for the charge $n$ monopoles to define a
 connection satisfying the Y-M equations of motion, while the
 corresponding projector does it. We conclude that at
 noncommutative level there is in general no equivalence relation
 between projectors and connections, as it happens in the
 classical case.

\section{Conclusions}

This work clarifies how to introduce topologically nontrivial
configurations on the fuzzy sphere. In particular it reveals the
mechanism with which the noncommutative topology can smoothly
extend the commutative one. In Ref. \cite{a21} we were able to
construct some noncommutative projectors having as entries the
elements of the fuzzy sphere algebra, and tending smoothly in the
$N \rightarrow \infty$ limit to the classical monopoles on the
sphere $S^2$. However at a physical level the classical monopoles
are also solutions of the equations of motion of the $Y-M$ action
on the sphere. Our first result is to identify a class of models
for which the noncommutative projectors \cite{a21} are solutions
of the corresponding equations of motion.

In the spirit of the Serre-Swan theorem we have then tried to
reconstruct the gauge connection or equivalently the matrix model
variables $X_i$ corresponding to these projectors. In this sense
we have reached a partially successful result because it is
possible to deconstruct only the projectors with negative charge
$-k$ ( $ p_{-k} , k > 0 $), but not those of positive charge $ p_k
, k > 0$.

In the $ p_{-k} ( k > 0 ) $ case, we have isolated the
corresponding matrix model solution, which results to be composed
in two steps:

i) with a re-scaling of the background solution, which is
necessary to require that the noncommutative topology has as a
classical smooth limit the commutative one;

ii) dressing with a quasi-unitary operator, that maps the
background, which is an irreducible representation of the $SU(2)$
Lie algebra, to a reducible representation. Therefore, at a level
of the matrix model, the classification of the topologically
nontrivial configurations is reduced to the classification of the
reducible representations of the $SU(2)$ Lie algebra, at least in
the fuzzy sphere case. The fact that we are able to find
connections only for negative charge projectors $ p_{-k} \ ( 0 < k
< N+1 ) $ is explained with the fact that in a $ (N+1) \times
(N+1) $ matrix variable $X_i$ we can insert a ( reducible )
representation with rank less than $(N+1)$, but not bigger than
$(N+1)$.

The case of positive charge projectors $ p_{k} \ ( k > 0 ) $ ,
where it is not possible to define a corresponding connection
satisfying the $Y-M$ equations of motion is a counterexample to an
eventual equivalence relation between projectors and connections,
as instead it happens in the classical case.


\begin{thebibliography}{999}

\bibitem{a17} J. Madore, " The Fuzzy sphere ", Class. Quantum Grav.
{\bf 9} (1992) 69.
\bibitem{a21} P. Valtancoli, " Projectors for the fuzzy sphere "
Mod.Phys.Lett. {\bf A16} (2001) 639, hep-th/0101189.
\bibitem{a18} S. Baez, A.P. Balachandran, B. Ydri, S. Vaidya, " Monopoles
and Solitons in fuzzy physics ", Comm. Math. Phys. {\bf 208}
(2000) 787,hep-th/9811169.
\bibitem{a19} A.P. Balachandran and S. Vaidya, " Instantons
and chiral anomaly in fuzzy physics ", hep-th/9910129.
\bibitem{a20} A. P. Balachandran, X.Martin and D. O' Connor, "
Fuzzy actions and their continuum limits ", hep-th/0007030.
\bibitem{a1} M.Dubois-Violette, Y. Georgelin, " Gauge theory in terms of
projector valued fields ", Phys. Lett. B {\bf 82} 251, 1979.
\bibitem{a2} T. Eguchi, P. B. Gilkey, A. J. Hanson, " Gravitation, gauge
theories and differential geometry ", Phys.Rept. {\bf 66}
213,1980.
\bibitem{a22} P. Valtancoli, " Stability of the fuzzy sphere solution from matrix
model", Int. J. Mod. Phys. {\bf A18} (2003), 967; hep-th/0206075
\bibitem{a23} P. Valtancoli, " Solitons for the fuzzy sphere from
matrix model",  Int. J. Mod. Phys. {\bf A18} (2003), 1107;
hep-th/0209117.
\bibitem{a6} S. Iso, Y. Kimura, K. Tanaka and K. Wakatsuki,
" Noncommutative gauge theory on fuzzy sphere from Matrix model ",
Nucl. Phys. {\bf B604} (2001) 121, hep-th/0101102.
\bibitem{a7} R.C. Myers, "Dielectric branes", JHEP {\bf 9912}
(1999) 022, hep-th/9910053.
\bibitem{a8} Y. Kimura, " Noncommutative gauge theories on fuzzy
sphere and fuzzy torus from Matrix model ", Prog. Theor. Phys.
{\bf 106} (2001) 445, hep-th/0103192.
\bibitem{a9} D. Berenstein, J. M. Maldacena, H. Nastase,
" Strings in flat space and PP waves from N=4 Superyang-mills " ,
JHEP {\bf 0204} (2002) 013, hep-th/0202021.
\bibitem{a10} J. Ambjorn, Y.M. Makeenko, J. Nishimura and R. J.
Szabo, " Finite N Matrix models of noncommutative gauge theory ",
JHEP {\bf 9911} (1999) 029, hep-th/9911041.
\bibitem{a11} J. Ambjorn, Y.M. Makeenko, J. Nishimura and R. J.
Szabo, " Non perturbative Dynamics of noncommutative gauge theory
", Phys. Lett. {\bf B480} (2000) 399.
\bibitem{a12} J. Ambjorn, Y.M. Makeenko, J. Nishimura and R. J.
Szabo, " Lattice gauge fields and discrete noncommutative
Yang-Mills Theory ", JHEP {\bf 0005} (2000) 023, hep-th/0004147.
\bibitem{a13} H. Aoki, S. Iso, H, Hawai, Y. Kitazawa, T. Tada and A.
Tsuchiya, " IIB matrix model ", Prog. Theor. Phys. Suppl. {\bf
134} (1999) 47, hep-th/9908038.
\bibitem{a14} H. Aoki, S. Iso, H. Hawai, Y. Kitazawa and T. Tada,
" Space-time structures from IIB Matrix model ", Prog. Theor.
Phys. {\bf 99} ( 1998) 713, hep-th/9802085.
\bibitem{a15} M. Li, " Strings from IIB matrices ", Nucl. Phys. {\bf
B499}(1997) 149, hep-th/9612222.
\bibitem{a16} H. Aoki, N. Ishibashi, S. Iso, H. Kawai, Y. Kitazawa
and T. Tada, " Noncommutative Tang-Mills in IIB Matrix model ",
Nucl. Phys. {\bf B565} (2000) 176, hep-th/9908141.
\bibitem{a3} G. Landi, " Projective modules of finite type and monopoles over
$S^2$ " , J.Geom.Phys. {\bf 37} 47, 2001, math-ph/9905014.
\bibitem{a24} P. Valtancoli, " Projective modules over the fuzzy four sphere",
Mod. Phys. Lett. {\bf A17} (2002) 2189; hep-th/0210166.
\end{thebibliography}
\end{document}